# Modeling and Simulation of Spark Streaming

Jia-Chun Lin, Ingrid Chieh Yu, and Einar Broch Johnsen
Department of Informatics, University of Oslo
Gaustadallèen 23 B, Oslo, N-0373, Norway
{kellylin, ingridcy, einarj}@ifi.uio.no

Ming-Chang Lee
Department of Communication Systems, Simula Research Laboratory
Martin Linges vei 25, Fornebu, 1364, Norway
mclee@simula.no

September 10, 2018

# Modeling and Simulation of Spark Streaming

Jia-Chun Lin
*Department of Informatics*
*University of Oslo*
Oslo, Norway
kellylin@ifi.uio.no

Ming-Chang Lee
*Department of Communication Systems*
*Simula Research Laboratory*
Oslo, Norway
mclee@simula.no

Ingrid Chieh Yu
*Department of Informatics*
*University of Oslo*
Oslo, Norway
ingridcy@ifi.uio.no

Einar Broch Johnsen
*Department of Informatics*
*University of Oslo*
Oslo, Norway
einarj@ifi.uio.no

*Abstract*—As more and more devices connect to Internet of Things, unbounded streams of data will be generated, which have to be processed "on the fly" in order to trigger automated actions and deliver real-time services. Spark Streaming is a popular real-time stream processing framework. To make efficient use of Spark Streaming and achieve stable stream processing, it requires a careful interplay between different parameter configurations. Mistakes may lead to significant resource over-provisioning and bad performance. To alleviate such issues, this paper develops an executable and configurable model named SSP (stands for Spark Streaming Processing) to model and simulate Spark Streaming. SSP is written in ABS, which is a formal, executable, and object-oriented language for modeling distributed systems by means of concurrent object groups. SSP allows users to rapidly evaluate and compare different parameter configurations without deploying their applications on a cluster/cloud. The simulation results show that SSP is able to mimic Spark Streaming in different scenarios.

## I. INTRODUCTION

Cyber-Physical Systems and Internet of Things (IoT) collect data from physical objects to improve quality, reduce risk, or increase profits in a range of application domains such as environmental monitoring, traffic control, predictive maintenance in the process industry, energy management, and healthcare, as well as smart infrastructure, buildings, and cities. According to Gartner [1] and ABI Research [2], we can expect to be surrounded by 26–30 billion IoT devices by the year 2020, producing enormous streams of data about us and the environment we live in.

Such hug amount of data stream requires real-time stream processing solutions that can process the data on the fly. During the past few years, different processing frameworks have been introduced, including Spark Streaming [3], AWS Kinesis Streams [4], and IBM InfoSphere Streams [5]. Among these, Spark Streaming attracts great attentions due to its scalability, efficiency, resilience, fault tolerance, and compatibility with several different cluster/cloud platforms, including Apache Hadoop YARN, Amazon EC2, and Mesos. However, achieving stable and efficient processing for a Spark streaming application requires a careful interplay between the configurations of the underlying stream processing framework and the streaming application itself. Making inappropriate configuration decisions may result in resource over-provisioning or poor performance.

In this paper, we present SSP, which is a highly configurable and executable model of Spark Streaming written in ABS [6, 7], a formal, executable language for modeling distributed systems and deployed virtualized software by means of concurrent object groups. SSP captures the key components of the Spark stream processing framework. It allows users to not only specify the workflow of their streaming applications, but also to configure crucial parameters of the underlying stream processing framework.

To show the faithfulness of the proposed model, we compare the simulated performance of SSP with the observed performance of Spark Streaming by designing several scenarios. The main contributions of this paper are as follows:

- SSP is, to the best of our knowledge, the first executable and configurable model to simulate both Spark stream processing framework and Spark stream applications.

- SSP allows users to specify the workflow of their streaming applications and experiment with different configuration settings for the stream processing framework. By means of simulations, users can easily evaluate different deployment decisions and predict the possible performance results.

- SSP is faithful to Spark Streaming. Our simulation results suggest that SSP provides a satisfactory modeling accuracy and reflects the operations of Spark Streaming.

The rest of the paper is organized as follows: We provide background on Spark Streaming in Section 2. Section 3 surveys related work. In Section 4, we present how SSP models and simulates Spark Streaming. Section 5 discusses the simulation results, and Section 6 concludes the paper and discusses our future work.

## II. BACKGROUND

As part of Apache Spark tool suite [8], Spark Streaming [3] is introduced to offer high-throughput and fault-tolerant data stream processing by using the discretized stream (D-Streams) model [9] to structure stream processing as a series of stateless, deterministic batch computations on small time intervals [10]. Each Spark streaming application is a long-running and non-terminating stream processing service with a Spark driver as its master, which is responsible for initiating the reception of data streams, periodically dividing the received data into batches based on a predefined batch interval, and managing the processing of each batch. A batch may contain zero to many data items depending on the data arrival frequency and batch interval. In this paper, a batch with no data is called an empty batch. Otherwise, it is called a non-empty batch. In addition to the above duties, the Spark driver also controls the maximum number of jobs that can be executed concurrently based on the concurrentJobs parameter (with the default value one [11]).

In Spark Streaming, each action (such as print or save as a file) triggers a job. Therefore, each non-empty batch is processed by one or a sequence of jobs depending on the program logic of the application designer. Each job (which we call a non-empty job) consists of several stages that can be executed in sequential, in parallel, or both. The corresponding workflow is a directed acyclic graph. Most spark stream applications have only one job. For example, the non-empty job shown in Figure 1 comprises 4 stages. Clearly, Stages 2 and 3 can execute in parallel after Stage 1 has finished, and Stage 4 can execute after Stages 2 and 3 have completed. On the other hand, each empty batch is also processed by a job with a single dummy stage (which we call an empty job). Note that Spark Streaming also allows users to determine a block interval for further dividing a batch into smaller blocks. The default value is 200 ms [12].

Spark Streaming supports several options for deployments; it can be deployed on Spark standalone clusters, YARN, Mesos, or on Amazon EC2. A Spark driver can request a set of worker nodes (which may be physical machines, virtual machines, or containers) to start its streaming application and process each batch on these worker nodes according to the corresponding workflow. More specifically, each block processed by a stage is performed by an executor of a worker node.

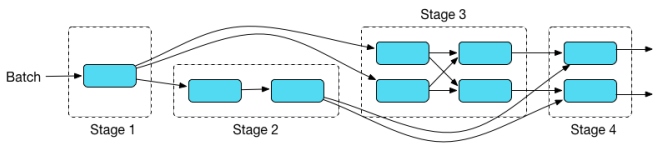

Fig. 1. An example of a non-empty job.

### III. RELATED WORK

During the last ten years, many models have been proposed for data stream modeling and stream processing modeling. Kroß and Krcmar [29] presented a modeling and simulation approach to predict the response time of Spark Streaming by using the Palladio component model. However, this approach does not allow users to configure the stream processing framework and their applications. Delcoigne et al. [13] proposed an analytical fluid model for the integration of streaming and data traffic in a multiservice network based on an M/G/1 processor sharing queue. SECRET [14] is a descriptive model for describing and predicting the behaviour of diverse stream processing engines. This model focuses on time-based windows and single-input query plans and gives an end-to-end view of the effects of different execution semantics. The model addresses different stream processing engines based on their syntax, their capabilities to support certain types of queries, and their execution models.

State Refinement [15] is a formal method for transforming a stream processing function into a state transition machine with input and output. In this method, states are the abstraction of input history and state transition function are derived using history abstractions. Persistent Turing Machines [16] endows classical Turing machines with a dynamic stream semantics by formalizing the intuitive notion of sequential interactive computation. Event Count Automata [17] is a state-based model for Stream Processing Systems by capturing the timing properties of data stream in terms of arrival and service pattern.

Some other work focuses on modeling stream queries. Babcock et al. [18] describe fundamental models and issues in developing a general-purpose data stream management system, especially related to stream query languages, requirements and challenges in query processing, and algorithmic issues. The authors extend standard SQL to allow the specification of sliding windows. Later, Kapitanova et al. [19] have proposed a formal specification language called MEDAL to model data stream queries and data admission control. This language is based on Petri nets and focuses on modeling different stream-processing features such as collaborative decision-making and temporal and spatial data dependencies.

In contrast to all the models mentioned above, this paper focuses on modeling and simulating Spark Streaming in both the processing framework and streaming applications. We capture the main features of Spark Streaming and address how processing capacities, data arrival patterns, and framework parameters affect the batch scheduling delay and batch processing time of streaming applications. Our model is executable and configurable and allows users to observe and compare the performance consequences of their streaming applications. To our knowledge, this is the first work on a formal executable model of Spark stream processing.

### IV. THE SSP MODEL

The SSP is written in ABS [6, 7], which is a formal, executable, object-oriented language for modeling distributed systems by means of concurrent object groups. Concurrent object groups execute in parallel and communicate by asynchronous method calls and futures. The reasons we adopted ABS are three-fold: 1) It is more flexible than an ad-hoc simulator, 2) it has parallel run-time support in Erlang [20], and 3) it natively supports the resources of CPU and memory and the notion of deployment components. Therefore, the task of modeling the performance of a single virtual machine instance can be easily finished in few lines of code.

In the following, we will introduce how SSP models both Spark stream applications the Spark stream processing framework.

*A. Modeling Spark Stream Applications*

In SSP, a batch is represented by a ABS data type with an identifier bID and an associated size bSize. In datatype definitions, the parameter names of a constructor become accessor functions, e.g., bID(Batch(1,5)) and bSize(Batch(1,5)) reduce to 1 and 5, respectively. To recognize empty batches, we define a Boolean function isEmptyBatch to check if the size of a batch is zero or not.

```
type BatchID = Int;
data Batch = Batch(BatchID bID, Int bSize);
def Bool isEmptyBatch(Batch batch)=(bSize(batch)==0);
```

In order not to complicate the modeling, we focus on batch-level modeling by assuming that the block interval equals to the batch interval specified by users. In addition, we focus on modeling streaming applications at stage level for one single job since in most cases streaming applications consist of only one job. This stage-level processing is modeled by means of two datatypes STJob and Stage. The former defines a non-empty job as a list of stages with their own identifiers; the latter defines a list of constraints constr for executing each stage.

```
type StageID = String;
data STJob = STJob(List<StageID> stages);
data Stage = Stage(StageID stID,List<StageID> constr);
```

For instance, the job workflow depicted in Figure 1 has four stages. Stage S1 can execute first since it has no dependencies, S2 and S3 can execute in parallel once S1 has finished, and S4 requires both S2 and S3 to be finished before it can execute. This workflow corresponds to the ABS term STJob[Stage(S1,Nil),Stage(S2,[S1]),Stage(S3,[S1]),Stage(S4,[S2,S3])]. To guarantee that all stages of a non-empty job are executed in accordance with their constraints, we define a Boolean function check as below:

```
def Bool check <A>(List<StageID> constr, List<StageID> fin)=
  case cs=constr {
    Nil => True;
    Cons(cs1,rest) => case element(cs1,fin)
    { True => checkConstraints(rest,fin);
      False => False;};};
```

For each unprocessed stage, this function recursively checks if all its constraints are included in fin, which is a list of completed stages of the job. A stage can be performed only when all its constraints have been resolved. SSP allows users to additionally define a cost approximation function costPerStage as below by assigning an execution cost expression $e_i$ depending on the batch size to each stage $s_i$ of the non-empty jobs (where $i = 1,2,...,n$), and a fixed cost expression e for the single stage of the empty job.

```
def Rat costPerStage(StageID stID, Int bSize) =
  case stID { "S1"=>e₁(bSize); "S2"=>e₂(bSize); ... ;
              "emptyJobstage"=>e;}
```

Notice that users can specify cost expressions at an appropriate level of precision using cost analysis (e.g., SACO [21]) or using an estimated or average execution time on batch size bSize.

*B. Modeling the Spark Stream Processing Framework*

Figure 2 illustrates the architecture of SSP, which consists of a main block for users to configure their streaming applications and the underlying Spark stream processing framework, a class SparkDriver to model the Spark driver of the streaming application, and a class Worker to model worker nodes for processing streaming applications. The following configuration parameters are available on SSP:

– the job workflow specification of a streaming application
– the execution cost for each stage of the job
– the number of worker nodes (denoted by num) used to run the application
– the resource specification for each worker node (denoted by rs)
– data inter-arrival pattern
– batch interval (i.e., bi)
– the maximum number of jobs allowed to execute concurrently (i.e., conJobs)

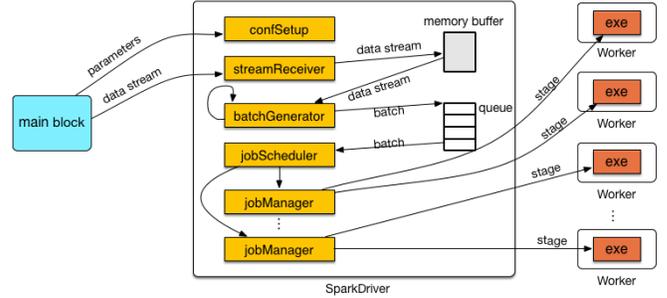

Fig. 2. The architecture of SSP.

The SparkDriver interface has the following five methods: confSetup, streamReceiver, batchGenerator, jobScheduler, and jobManager.

```
interface SparkDriver {
Bool confSetup(Int num, RSpec rs, Int bi, Int conJobs);
Unit streamReceiver(Int sizeStream);
Unit batchGenerator();
Unit jobScheduler();
Unit jobManager(Batch batch);}
```

When a user launches SSP, the method confSetup is invoked to create the requested set of num worker nodes with the requested resource specification rs. Each worker node is modeled by class Worker on an independent deployment component with the resource specification rs of type RSpec:

**data** RSpec = Res(Int cores, Rat speed, Int memory)

Class Worker offers a method exe. Due to the batch-level modeling, each stage of a streaming job is executed by method exe with a cost expression e. The stage execution time is determined by the cost expression e and the CPU processing speed of the worker node (i.e., the corresponding deployment component).

After all the worker nodes have been created, the method confSetup triggers methods streamReceiver, batchGenerator, and jobScheduler to start running the streaming application. Streams of data are sent to the method streamReceiver based on the pre-configured data inter-arrival pattern. Upon receiving data, method streamReceiver keeps it in the memory buffer of the Spark driver. The method batchGenerator periodically generates batches according to the pre-configured batch interval and inserts them into the queue. Method jobScheduler schedules all batches by separately creating a jobManager to manage one batch processing. For brevity, we here focus on describing methods batchGenerator, jobScheduler, and joManager.

**Method batchGenerator**

The method batchGenerator, as shown in Figure 3, is a non-terminating process to periodically generate batches based on batch interval bi. Using the await duration statement of ABS, this method suspends its execution and resumes again after the time interval bi. Each time, batchGenerator collects all received data in the memory buffer into a batch, inserts the batch into the queue, and empties the buffer to avoid reprocessing the same data. Each batch has a unique bID and has size bSize where bSize=DataSizeInBuffer, i.e., total size of the data in the memory buffer.

```
Unit batchGenerator (){
  Int bID = 1; Int bSize=0;
  while (True){
    await duration(bi,bi);
    bSize=DataSizeInBuffer;
    queue = appendright(queue,Batch(bID,bSize));
    DataSizeInBuffer = 0; bID = bID + 1;
}}
```
Fig. 3. The algorithm of method batchGenerator.

**Method jobScheduler**

Method jobScheduler is also a non-terminating process (see Figure 4). It uses the default first-in-first-out scheduling approach to schedule the execution of batches. If the total number of currently executing jobs is less than conJobs (i.e., runningJob<conJobs) and the queue has an unprocessed batch, jobScheduler creates a jobManager object to manage the processing of the head-of-queue batch by invoking method jobManager. The await statement suspends the active process in ABS and therefore allows SparkDriver to interleave the different stream processing activities in a flexible way.

```
Unit jobScheduler (){
  while (True){
    await (runningJob<conJobs);
    await (length(queue)>0);
    Batch batch = head(queue);
    this!jobManager(batch);
    queue=tail(queue);
    runningJob=runningJob+1;
}}
```
Fig. 4. The algorithm of method jobScheduler.

**Method jobManager**

Figure 5 illustrates the algorithm of method jobManager. Whenever a jobManager object is generated to manage the processing of a batch, it retrieves the corresponding workflow (i.e., jobStageList) depending on whether the batch is empty. As long as there are unfinished stages (i.e., length(fin) < totalStages), the jobManager checks the corresponding constraints to see if any of these stages can be executed immediately. When all the constraints of a stage s are resolved (i.e., check(constr(s),fin) equals to True), the jobManager attempts to find a worker node from the workerList and execute the stage on the worker node by invoking its exe method. When all stages of the job are finished, the batch processing is complete. In this case, the parameter runningJob decreases by one, which again allows the jobScheduler to schedule another batch for processing.

```
Unit jobManager (Batch batch){
  List<StageID> fin=Nil;
  List<Stage> jobStageList = case (isEmptyBatch(batch)){
    True => stages(emptyJob); False => stags(nonEmptyJob);};
  Int totalStages=length(jobStageList);
  while (length(fin)<totalStages){
    Stage s = head(jobStageList);
    Bool canItRun = check(constr(s),fin);
    if (canItRun==True){
      await (length(workerList)>0);
      Worker wr= head(workerList);
      workerList=tail(workList);
      Bool res=await wr!exe(costPerStage(stID(s),bSize(batch)));
      if (res){
        fin = appendright(fin,stID(s));
        workerList = appendright(workerList,wr);
        jobStageList=tail(jobStageList);}}
    else{
      jobStageList = appendright(tail(jobStageList),s);}
    await duration(1,1);
}
if (length(fin)==totalStages){ //The job is finished.
  runningJob=runningJob-1;
  ... //Record the time info of the job.}}
```
Fig. 5. The algorithm of method jobManager.

## V. RESULT COMPARISON

To validate the faithfulness of SSP, we compared the simulation results of SSP with the execution results of Spark Streaming deployed on a YARN cluster running Hadoop 2.2.0 [22] and Spark 1.5.1 [23]. The cluster has 30 virtual machines acting as worker nodes. Each virtual machine runs Ubuntu 12.04 with 2 virtual cores of Intel Xeon E5-2620 2GHz CPU and 2 GB of memory. To achieve a fair comparison, we similarly configured the SSP model to have 30 worker nodes; each with 2 CPU cores and 2 GB of memory.

We chose a well-established Spark Streaming benchmark called JavaNetworkWordCount [24] as an example application. This application consists of one job, which has two sequential stages to count the occurrences of each word in data streams. We employed Netcat [25] to continuously send data from the Wikipedia website of Apache Spark [26] to the application. The size of each data item is 1 KB. The data is sent in a dynamic inter-arrival pattern following an exponential distribution [27] with the average time of 1.96 sec and the standard deviation of 1.768 sec. We executed JavaNetworkWordCount on Spark Streaming, and obtained the results that the time for processing an empty batch is 0.1 sec, the time for processing stage 1 of a non-empty batch varies between 3.1 to 3.4 sec, and the time for processing stage 2 of a non-empty batch is 0.1 sec. To show modeling effects, we configured our model by enlarging each of the above time values by 10 times. This process is called *normalization*. In the following, we evaluate if SSP captures the following key properties of Spark Streaming:

- Property 1: Batches are generated based on the pre-configured bi.
- Property 2: If the data inter-arrival time is shorter than bi, the generated batches will be all non-empty. Otherwise, some batches might be empty.

– Property 3: The scheduling for batch processing is based on the FIFO order.

The following two performance metrics are also used for comparison:

– Scheduling delay: The time a batch waits for scheduling.
– Processing time: The total time to process a batch.

In Spark Streaming, a streaming application is stable if each of its batches can be scheduled immediately. For stability, the parameters conJobs and bi are highly influential, so we designed two scenarios to configure both parameters.

### A. Scenario 1: No Concurrent Job Processing

In Scenario 1, at most one job is allowed to be processed at any time and a batch is periodically generated every two seconds, implying that conJobs=1 and bi=2 sec. Figure 6 shows the processing start time of each batch in both SSP and Spark Streaming. We can see that in both environments, the batch scheduling always follows the FIFO order. A batch with a smaller batch ID always has an earlier processing start time, implying that SSP captures property 3. Figure 7 shows the time period between batch generation in both SSP and Spark Streaming. We can see that every generation interval between two consecutive batches is always two seconds, which is exactly equal to the value of bi in Scenario 1. It means that SSP captures Property 1, i.e., batches are periodically generated based on the pre-configured bi.

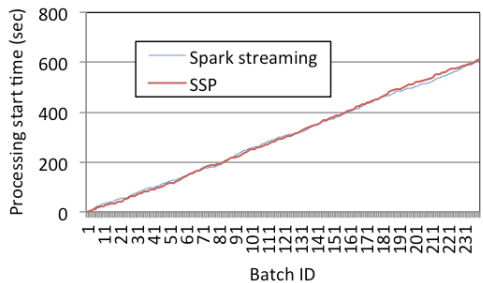

Fig. 6. The batch processing start time of both Spark Streaming and SSP under Scenario 1.

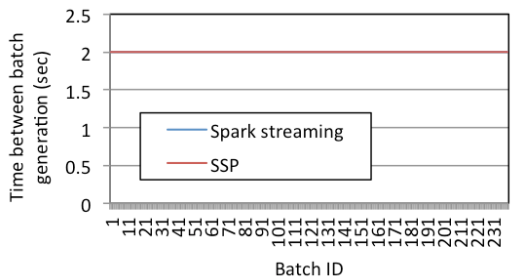

Fig. 7. The time between batch generation in both Spark Streaming and SSP under Scenario 1.

Figure 8 shows that the batch scheduling delay in Spark Streaming keeps increasing as more batches are generated, implying that Spark Streaming was unable to immediately allocate containers to process these batches. The possible reasons are twofold: First, the value of conJobs might be insufficient, so other batches need to wait in the queue. Second, setting bi to two seconds may be too short. A shorter bi implies that batches are generated more frequently, so more worker nodes are required to process these batches. Despite the poor performance, it is clear that the batch scheduling delay in SSP follows the same trend because SSP captures both the batch processing and job concurrency aspects of Spark Streaming.

Figure 9 illustrates the batch processing time in both SSP and Spark Streaming. Because data arrived every 1.96 sec in average and the standard deviation was 1.768 sec, data might sometimes arrive frequently and sometimes more seldom. Due to the fact that bi is 2 sec, a lot of generated batches were empty. This explains why the processing time of batches in Spark Streaming fluctuated between 4 sec and 0 sec. This phenomenon also occurs in SSP since it models both empty and non-empty batch processing. The results show that SSP captures Property 2, i.e., empty batches will be generated when data inter-arrival time is longer than bi.

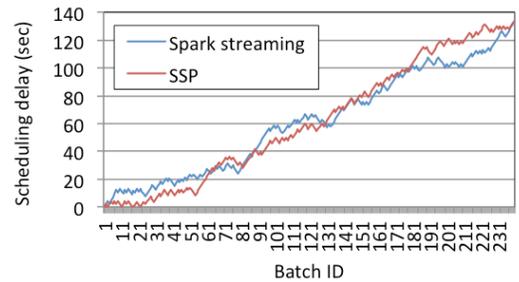

Fig. 8. The batch scheduling delay of both Spark Streaming and SSP under Scenario 1.

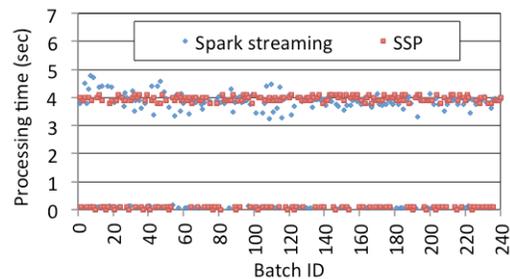

Fig. 9. The batch processing time of both Spark Streaming and SSP under Scenario 1.

### B. Scenario 2: Concurrent Job Processing

It is clear that the configuration of Scenario 1 is inappropriate for JavaNetworkWordCount under the dynamic data inter-arrival pattern. Hence, we increased bi from 2 to 4 sec while leaving conJobs at 1. However, this reconfiguration is still unable to improve the stability of the application. Therefore, we kept bi as 4 sec and further increased conJobs from 1 to 15. This setting is called Scenario 2, which we used to configure SSP.

The results of batch processing start time and batch generation intervals for Scenario 2 are shown in Figures 10 and 11, respectively. These figures suggest that SSP is able to mimic how Spark Streaming schedules batches since the time curve of SSP almost overlaps with that of Spark Streaming.

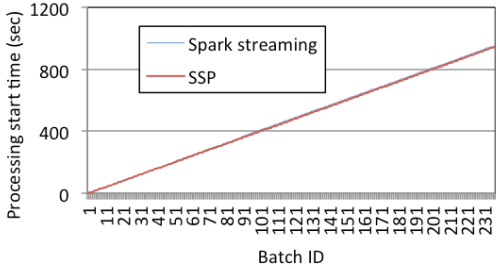

Fig. 10. The batch processing start time of both Spark Streaming and SSP under Scenario 2.

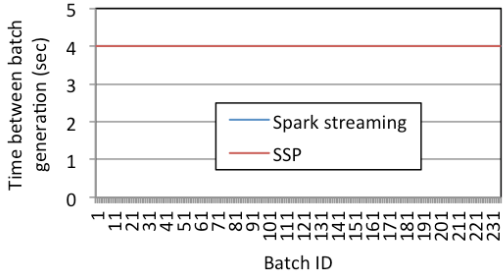

Fig. 11. The time between batch generation in both Spark Streaming and SSP under Scenario 2.

Figure 12 shows that the application in Spark Streaming was more stable in Scenario 2 compared with that in Scenario 1. We can see that the batch scheduling delays were significantly reduced and in fact were close to zero second. This means that each batch could be scheduled almost immediately. This phenomenon is also captured and reflected by SSP. Figure 13 shows that the number of empty batches in Spark Streaming in Scenario 2 is far lower than in Scenario 1 (please compare Figure 13 with Figure 9). The main reason is that the value of $b_i$ in Scenario 2 was double of the value in Scenario 1, which means that the probability that no data arrives within the batch interval is reduced. We see that our SSP model also captures this change.

Based on the above results, we conclude that SSP indeed captures the key properties of Spark Streaming discussed above and it provides a good approximation of the performance of Spark Streaming under the dynamic data traffic pattern for both Scenario 1 and Scenario 2. Users can easily model their streaming applications in SSP and compare how different parameter configurations affect the performance of their applications before these applications are deployed by means of simulations.

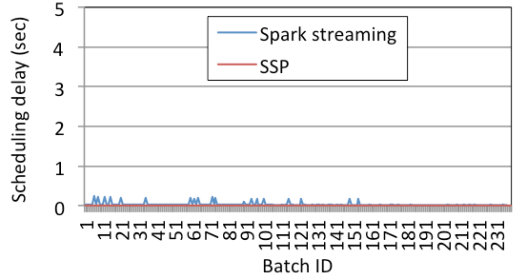

Fig. 12. The batch scheduling delay of both Spark Streaming and SSP under Scenario 2.

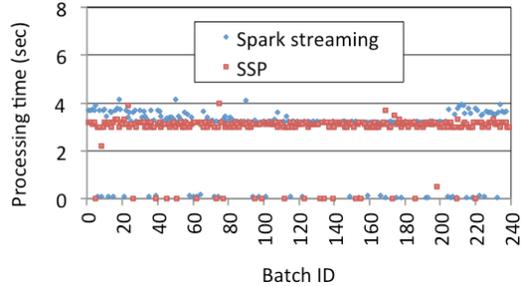

Fig. 13. The batch processing time of both Spark Streaming and SSP under Scenario 2.

## VI. CONCLUSIONS AND FUTURE WORK

In this paper, we have presented SSP for modeling Spark Streaming. The proposed model enables users to configure the processing framework of Spark Streaming and adapt it to their streaming application settings, including streaming workflow and stage execution cost. We believe that it is crucial to show that the proposed model can faithfully reflect Spark Streaming once the model has been configured. To validate the proposed SSP model, we have compared it with Spark Streaming running on a YARN cluster. The results show that 1) the SSP model captures the key properties of Spark Streaming, including batch generation, empty batch processing, non-empty batch processing, and batch scheduling; 2) the model provides a good approximation of Spark Streaming in terms of the batch scheduling delay and batch processing time; and 3) the model enables users to predict the performance of their streaming applications on Spark stream processing framework with different configuration settings during the modeling phase and thereby to determine an appropriate configuration setting.

In future work, we plan to extend the SSP model by considering spark streaming applications with a sequence of jobs and meanwhile taking the block-level modeling into account. The model will also be integrated with ABS-YARN [28], which is an executable model framework for modeling and simulating YARN, to allow users to further configure the resources of Spark stream processing framework for their streaming applications. In addition, we would like to extend the model by taking the reliability of the underlying cluster/cloud infrastructure into account, such as modeling the failures of worker nodes and network connections.

ACKNOWLEDGMENT

This work was supported by the European project HyVar (grant agreement H2020-644298). The authors would like to thank the anonymous reviewers for their valuable comments to improve the quality of this paper.